\documentclass[aps,prb,superscriptaddress]{revtex4-2}

\usepackage[utf8]{inputenc}
\usepackage{bm}
\usepackage{amsmath}
\usepackage{amssymb}
\usepackage{caption}
\usepackage{graphicx}
\graphicspath{{img_1/}}
\usepackage{color}

\usepackage{caption}
\usepackage{subcaption}

\input{epsf}
\usepackage[english]{babel}

\begin{document}

\title{Nanomechanics driven by the superconducting proximity effect}

\author{O.M.~Bahrova}
\email{bahrova@ilt.kharkov.ua}
\affiliation{Center for Theoretical Physics of Complex Systems, Institute for Basic Science, Daejeon, 34126, Republic of Korea}
\affiliation{B. Verkin Institute for Low Temperature Physics and
Engineering of the National Academy of Sciences of Ukraine, 47 Nauky Ave., Kharkiv 61103, Ukraine}

\author{S.I.~Kulinich}
\affiliation{B. Verkin Institute for Low Temperature Physics and
Engineering of the National Academy of Sciences of Ukraine, 47 Nauky Ave., Kharkiv 61103, Ukraine}

\author{L.Y.~Gorelik}
\affiliation{Department of Physics, Chalmers University of
Technology, SE-412 96 G{\" o}teborg, Sweden}

\author{R.I.~Shekhter}
\affiliation{Department of Physics, University of Gothenburg, SE-412 96 G{\" o}teborg, Sweden}

\author{H.C.~Park}
\email{hc2725@gmail.com}
\affiliation{Center for Theoretical Physics of Complex Systems, Institute for Basic Science, Daejeon, 34126, Republic of Korea}

\date{\today}

\begin{abstract}

We consider a nanoelectromechanical weak link composed of a
carbon nanotube suspended above a trench in a normal metal electrode and positioned in a gap
between two superconducting leads. The nanotube is treated as a movable
single-level quantum dot in which the position-dependent
superconducting order parameter is induced as a result of Cooper
pair tunneling. We show that in such a system,
self-sustained bending vibrations can emerge if a bias voltage is
applied between normal and superconducting electrodes. The occurrence of this effect crucially depends on the
direction of the bias voltage and the relative position of the quantum dot
level. We also demonstrate that the nanotube vibrations strongly affect
the dc current through the system, a characteristic that can be used for the
direct experimental observation of the predicted phenomenon.

\end{abstract}
\maketitle

\section{Introduction}

Nanoelectromechanical systems (NEMS) provide a promising platform for investigations into the
quantum mechanical interplay between mechanical and electronic subsystems~\cite{Cleland,Ekinci}.
One of the most important phenomena providing the foundation of NEMS functionality
is the generation of self-sustained mechanical oscillations by a dc flow~\cite{firstsh, BlanterNazarov, fedorets1, kulinich, ilinskaya, Anton}.
This effect is itself an interesting problem from a fundamental point of view, opening new
possibilities for mass and force sensing~\cite{zettl,braakman}, while its underlying physical processes show potential
applications for mechanical cooling~\cite{urgell}. Self-sustained mechanical oscillations were first
observed in a carbon nanotube (CNT)-based transistor~\cite{zant}, with further studies later verifying
their transport signatures~\cite{schmid1,schmid2,zettl2}. Recently, the experimental observation of self-driven
oscillations of a CNT-based quantum dot in the Coulomb blockade regime has been reported~\cite{willick}.

Superconducting elements incorporated into NEMS extend the horizon of this phenomenon,
namely through the effects of superconducting phase coherence; see, for example, the following reviews~\cite{review, meden}. Interplay between electromechanical effects and phase coherence gives new and
unusual properties to a number of normal metal/superconducting hybrid junctions~\cite{baranski1, domanski2, konig}.
In particular, it has recently been shown that in a normal metal--suspended CNT--superconductor
transistor, Andreev reflection~\cite{andreev,kulik} may give rise to a cooling of the mechanical subsystem~\cite{stadler} or
generate a single-atom lasing effect~\cite{rastelli2} if certain conditions are fulfilled.

The mechanical functionality of NEMS is to a large extent determined by the physical principles
underlying the interaction between the electronic and mechanical subsystems. In all studies
mentioned above, this interaction was due to the \textit{localization} of the charge~\cite{BlanterNazarov,fedorets1,Anton} or spin~\cite{kulinich, ilinskaya} carried by
electrons in the movable part of the system. In the present paper, we consider a fundamentally new type
of electromechanical coupling based on the \textit{quantum delocalization} of Cooper pairs. We
demonstrate that such coupling can promote a self-saturated mechanical instability resulting in
the generation of \textit{self-sustained} mechanical oscillations. It is also shown that these oscillations significantly
increase the average current through the system, making it possible to carry out direct
experimental detection.

\section{Model and Dynamics}

\begin{figure}
\includegraphics[width=.45\columnwidth]{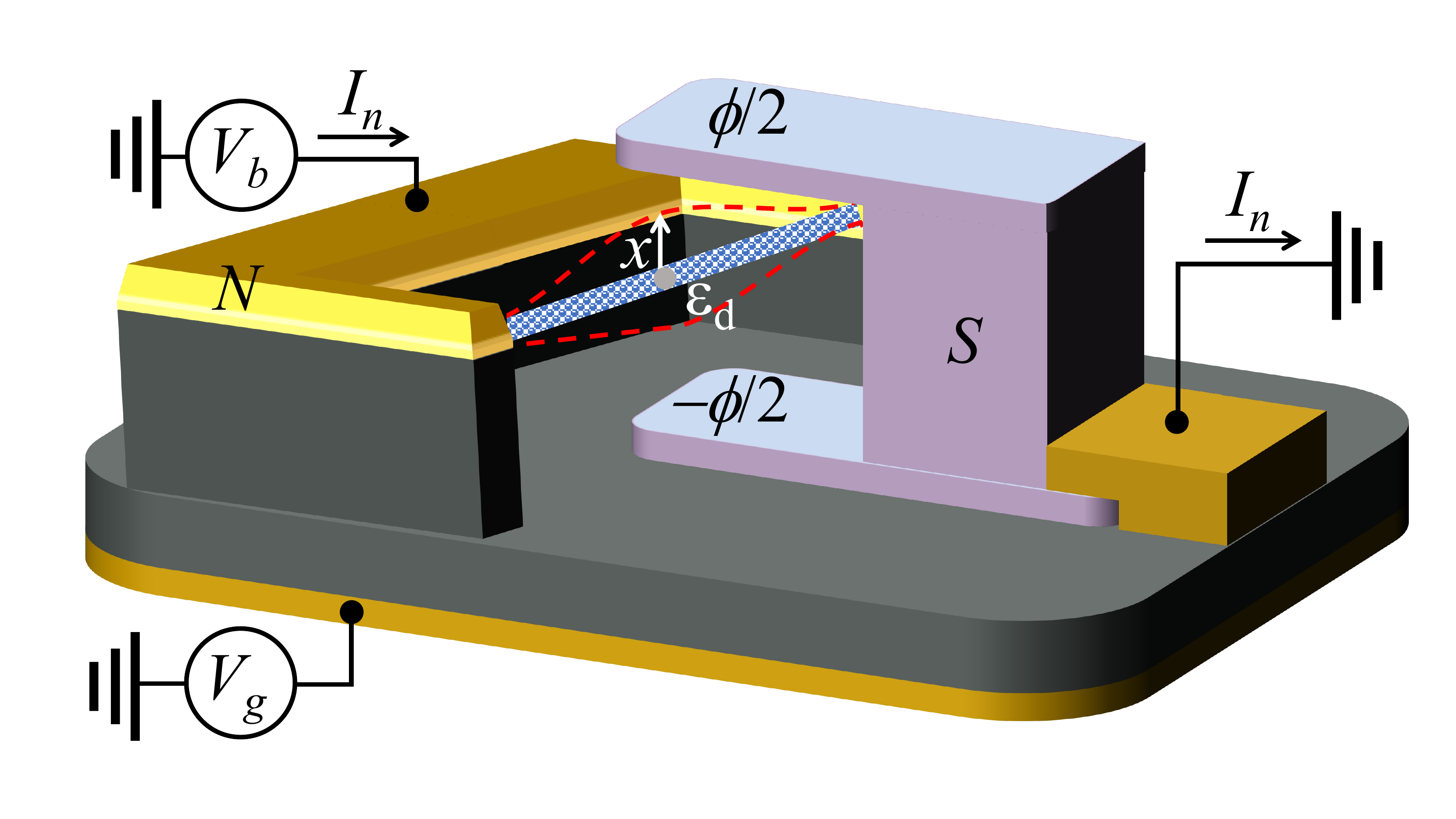}
\caption{Schematic illustration of the nanoelectromechanical device under consideration. A carbon nanotube (CNT) is suspended in a gap between two edges of a normal electrode ($N$) and tunnel-coupled to it. The electronic energy levels of the CNT are tuned such that only one energy level with energy $\varepsilon_d$, which is well separated from the other levels, is considered. Bending of the CNT in the $x$ direction between two superconducting leads ($S$) affects the values of the tunneling barriers between them. Bias voltage $V_b$ is applied to the normal electrode.
}\label{fig:fig1}
\end{figure}

A sketch of the NEMS investigated in this paper is presented in Fig.~\ref{fig:fig1}. A single-walled CNT is suspended above a trench in a bulk normal metal electrode
biased by a constant voltage $V_b$. Two superconducting
leads with the superconducting phase difference $\phi$ are positioned
near the middle of the nanotube in such a way that the bending of
the nanotube moves it closer to one electrode and further
away from the other. The distance between the quantized electronic
levels inside the nanotube is much greater than the other energy
parameters, allowing one to consider the nanotube as a single-level quantum dot (QD). The bending dynamics of the CNT are reduced to the
dynamics of the fundamental flexural mode. We suppose that the
amplitude of this mode, $x$, is larger than the amplitude of
zero-point oscillations, and will consider it as a classical
mechanical oscillator with mass $m$ and frequency $\omega$. The
dynamics is described by Newton's equation,
\begin{equation}\label{xEq1}
    \ddot{x}+\omega^2 x=-\frac{1}{m}\text{Tr}\left\{\hat{\rho}\frac{\partial H(x)}{\partial x}\right\},
\end{equation}
where
\begin{equation}\label{H}
    H=H_d+H_l+H_t
\end{equation}
is the Hamiltonian of the electronic subsystem. The first term
$H_d$ represents the single-level QD,
\begin{equation}\label{Hd}
    H_d=\sum_{\sigma}\varepsilon_d d^\dag_\sigma d_\sigma,
\end{equation}
where $d_\sigma^\dag (d_\sigma)$ is the creation (annihilation) operator of an electron with spin projection $\sigma=\uparrow,\downarrow$ on the dot.
The Hamiltonian $H_l=H_l^n+H_l^s$ describes the normal and superconducting leads, respectively, with
\begin{eqnarray}\label{Hl}
   && H_l^n=\sum_{k\sigma }(\varepsilon_k-eV_b)a_{k \sigma}^\dag a_{k\sigma},\\
    && H_l^s=\sum_{k j\sigma}\left(\varepsilon_{k}c^\dag_{kj\sigma}c_{kj\sigma}-\Delta_{s} (\text{e}^{\imath\phi_j } c^\dag_{kj\uparrow}c^\dag_{-kj\downarrow}+\text{H.c.})\right).
\end{eqnarray}
Here, $a^\dag_{k\sigma} (a_{k\sigma})$, and $c^\dag_{kj\sigma}
(c_{k\sigma})$ are the creation (annihilation) operators of an electron
with quantum number $k$ and spin projection $\sigma$ in the normal
and superconducting $j=1,2$ leads, respectively, and
$\Delta_{s}\text{e}^{\imath\phi_j}$ is the superconducting order
parameter (in the $j$ electrode). Note that the energies
$\varepsilon_{d},\varepsilon_{k}$ are counted from the Fermi energy of
the superconductors. In what follows, we set $\phi_{1}=-\phi_{2}=\phi/2$.

The Hamiltonian $H_t=H_t^n+H_t^s$ describes the tunneling of
electrons between the dot and the leads, where
  \begin{eqnarray}\label{Ht}
  && H_t^n=\sum_{k\sigma}t^n_0 (a_{k\sigma}^\dag d_\sigma +\text{H.c.}), \\
 &&  H_t^s=\sum_{k j\sigma}t^s_{ j}(x) (c^\dag_{kj\sigma}d_\sigma +\text{H.c.}).\label{Hts}
\end{eqnarray}
The position-dependent superconducting tunneling amplitude
$t^s_{1(2)}(x)=t^s_0 \text{e}^{(-1)^j (x+a)/2\lambda}$, where
$2\lambda$ is the characteristic tunneling length and $a$ is a
parameter for asymmetry. For a typical CNT-based nanomechanical
resonator, $2\lambda\sim 0.5$ nm~\cite{GatecontrolledPE}. We concentrate our attention on the symmetric case
$a=0$ and leave the asymmetric one for discussion elsewhere.

The time evolution of the electronic density matrix $\hat{\rho}$ is described by the Liouville–von Neumann equation ($\hbar=1$),
  \begin{equation}\label{LvN}
      \imath \partial_t \hat{\rho}=[H,\hat{\rho}],
  \end{equation}
which together with Eq.~(\ref{xEq1}) forms a closed system of
equations that describe the nanoelectromechanics of our system. In
this paper, we restrict ourselves to the case $\Delta_{s}\gg|eV_b|
\gg \Delta_d \sim \Gamma_n$, where $\Delta_d=(2\pi)\nu_{s}|t^{s}_0|^2$ and $\Gamma_n=(2\pi)\nu_{n}|t^{n}_0|^2$, with $\nu_{s(n)}$ the density of
states in the superconducting (normal) electrode.

To describe the electronic dynamics of the QD, we use the reduced density
matrix approximation in which the full density matrix of the system
is factorized to the tensor product of the equilibrium density
matrices of the normal and superconducting leads and the density matrix of the dot as
$\hat{\rho}=\hat{\rho}_n\otimes\hat{\rho}_s\otimes\hat{\rho}_d$.
Using the standard procedure, one can trace out the degrees of
freedom of the leads
 and obtain the following equation for the reduced density matrix $\hat{\rho}_d$~\cite{Anton} (in the deep subgap regime $\Delta_{s}\rightarrow
\infty$),
\begin{equation}\label{EDM}
  \partial_t \hat{\rho}_d=-\imath \left[H_d^{eff},\hat{\rho}_d \right]+\mathcal{L}_n\{\hat{\rho}_d\},
\end{equation}
where
\begin{equation}\label{Hdeff}
    H^{eff}_d=H_d+\Delta_{d}(x,\phi)d_{\downarrow}d_{\uparrow} +\Delta_{d}^* (x,\phi) d_\uparrow^\dag d_\downarrow^\dag,
\end{equation}
$$\Delta_{d}(x,\phi)=\Delta'(x,\phi)+i\Delta''(x,\phi)= \Delta_d \cosh(x/\lambda+i\phi/2).$$
Above, $\Delta_d(x,\phi)$ is the off-diagonal order parameter induced by the superconducting
proximity effect~\cite{rozhkov, stadler}, and $\Delta^{\prime,\prime\prime}(x,\phi)$ are real functions. The Lindbladian term in
Eq.~(\ref{EDM}) reflects the incoherent electron exchange
between the normal lead and QD. The latter in the high bias voltage
regime, $|eV_b|\gg \varepsilon_0, k_BT$, takes the form
\begin{equation}\label{Ln}
\mathcal{L}_n\{\hat{\rho}_d\}=\Gamma_n\sum\limits_\sigma
\begin{cases}
2 d^\dag_\sigma\hat{\rho}_d d_\sigma-\left\{ d_\sigma d_\sigma^\dag ,\hat{\rho}_d \right\} , & \kappa =+1; \\
2 d_\sigma\hat{\rho}_d d^\dag_\sigma-\left\{ d^\dag_\sigma d_\sigma ,\hat{\rho}_d \right\} , & \kappa=-1;
\end{cases}
\end{equation}
where $\kappa=\text{sgn}(eV_b)$.

\begin{figure}
    \centering
    \begin{subfigure}[t]{0.4\columnwidth}
        \centering
        \includegraphics[width=\columnwidth, keepaspectratio]
        {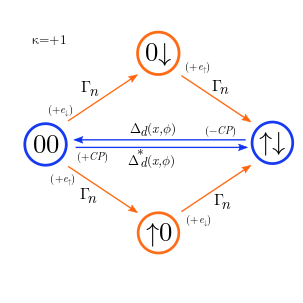}
        \caption{}\label{fig:subfig2a}
    \end{subfigure}
    \quad
    \begin{subfigure}[t]{0.4\columnwidth}
        \centering
        \includegraphics[width=\columnwidth, keepaspectratio]
        {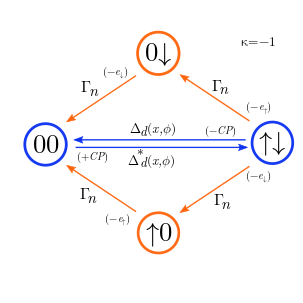}
        \caption{}\label{fig:subfig2b}
    \end{subfigure}
    \caption{Diagrams representing the transitions between electronic states in the quantum dot. The single-electron states change due to transitions from the empty to the single-occupied QD and then from the single-occupied to the double-occupied one (indicated by orange arrows). In the high bias voltage regime, the tunneling of electrons (a) or holes (b) with spin $\downarrow$ or $\uparrow$ is allowed only from the normal lead to the dot and forbidden in the opposite direction. Transitions between the empty and double-occupied quantum dot are due to coupling with the superconducting leads (indicated by blue arrows). }\label{fig:fig2}
\end{figure}

Figure~\ref{fig:fig2} represents the electronic dynamics on the dot for
$\kappa=\pm1$. From Fig.~\ref{fig:fig2}, one can see that not all electron processes are allowed due to the parameter scales in this work. In the subgap regime, single-electron transitions between the dot and the superconducting leads are prohibited, and thus only an exchange of Cooper pairs occurs. Moreover, because of the high bias voltage, single-electron tunneling between the dot and the normal leads is enabled exclusively in one direction (from the lead to the dot, see Fig.~\ref{fig:subfig2a}, or vice-versa, Fig.~\ref{fig:subfig2b}), establishing that our model is electron-hole symmetric.
As a consequence, the QD density matrix $\hat{\rho}_d$ acts in the Hilbert  space $\mathcal{H}_{4}$, which may be presented as a direct sum of
two $\mathcal{H}_{2}$ spaces via $\mathcal{H}_{4}=\mathcal{H}_{e}\oplus\mathcal{H}_{CP}$ spanned over state vectors
$\vert\uparrow\rangle = d^\dag_\uparrow\vert 0\rangle$, $\vert\downarrow \rangle = d^\dag_\downarrow\vert 0\rangle$,
and $\vert 0\rangle$, $\vert 2\rangle = d^\dag_\uparrow d^\dag_\downarrow\vert 0\rangle$ (with $d_{\uparrow,\downarrow}\vert 0\rangle=0$).

The superselection rule, which forbids the superposition of states with integer and half-integer spins, allows us to present the density matrix $\hat{\rho}_d$
as a direct sum of two density matrices $\hat{\rho}_d=\hat{\rho}_e\oplus\hat{\rho}_{CP}$ acting in the
$\mathcal{H}_{2}$ Hilbert space spanned over state vectors $\vert\uparrow\rangle,\vert\downarrow\rangle$ and $\vert 0\rangle,\vert 2\rangle$, respectively.
 Moreover, taking into account spin-rotation symmetry, one can conclude that $\hat{\rho}_e$ should be proportional to the unit matrix, $\hat{\rho}_e =\rho_e\hat{I}$,
 while $\hat{\rho}_{CP}$ can be written in the form
$\hat{\rho}_{CP}=\frac{1}{2}R_{0}\hat{I}+\frac{1}{2}\sum_i R_i\sigma_i$, where $\sigma_i ,(i=1,2,3)$ are the Pauli matrices.

Then by introducing the dimensionless time $\omega t\rightarrow t$ and displacement $x/\lambda \rightarrow x$, and taking into account the normalization condition
 $\text{Tr}\hat{\rho}_d=1$, we get the following closed system of equations for $x(t)$ and $R_{i}(t)$,

\begin{equation}
   \ddot{x}+x=-\xi\left[\sinh{(x)}\cos{\left(\frac{\phi}{2}\right)}R_{1}-\cosh{(x)}\sin{\left(\frac{\phi}{2}\right)}R_{2} \right],\label{xEq}
\end{equation}
\begin{equation}
    \alpha\dot{\vec{R}}=\hat{L}\vec{R}- \kappa\vec{e}_3,\label{EqR}
\end{equation}

where $\vec{R}=(R_{1},R_{2},R_{3})^{T}$, $\vec{e_3}=(0,0,1)^T$, $\xi=\Delta_{d}/( m\lambda^{2}\omega^{2})$ is the
nanoelectromechanical coupling parameter, and $\alpha=\omega /(2\Gamma_n)$ is the adiabaticity parameter. For a typical CNT-based
NEMS, one can estimate $\xi\sim 10^{-3}\ll 1$~\cite{GatecontrolledPE,Q5mil}. The
matrix $\hat{L}$ is defined as follows,

\begin{equation}\label{matrixA}
\hat{L}(x)=
\begin{pmatrix}
-1 &  \tilde{\varepsilon}_d & -  \tilde{\Delta}'' (x,\phi) \\
-\tilde{\varepsilon}_d & -1 & -\tilde{\Delta}'(x,\phi) \\
\tilde{\Delta}''(x,\phi) & \tilde{\Delta}'(x,\phi) & -1
\end{pmatrix},
\end{equation}
where $\tilde{\varepsilon}_d\equiv\varepsilon_d/\Gamma_n, \tilde{\Delta}_d\equiv\Delta_d/\Gamma_n$.

The system of Eqs.~(\ref{xEq}) and (\ref{EqR}) has an obvious static
solution $x_{st}=0+\mathcal{O}(\xi)$,
$\vec{R}_{st}=\kappa
L^{-1}(0)\vec{e}_{3}+\mathcal{O}(\xi)\vec{R}^{(1)}$, here
$\|\vec{R}^{(1)}\|=1$. The stability of this solution can then be
investigated in a standard way, see for example Ref.~\cite{ilinskaya2}.
However, to simplify this procedure, we will consider the adiabatic
case when $\alpha\ll 1$, which corresponds to a
typical experimental situation~\cite{willick} and reduces the
problem to one that allows the use of Poincare analysis. More specifically, this
inequality allows one to find a solution of Eq.~(\ref{EqR}) to the
accuracy $\alpha$,
\begin{equation}\label{Rexpand}
    \vec{R}(x,t)= \kappa L^{-1}(x(t))(1+\alpha
\dot{x}\partial_{x}L^{-1}(x(t))+\mathcal{O}(\alpha^{2}))\vec{e}_{3},
\end{equation}
and then substituting this solution into Eq.~(\ref{xEq}) gives
(to accuracy $\alpha$) the following nonlinear differential equation
for $x(t)$,
\begin{equation}\label{xEqsol}
    \ddot{x}-\eta(x,\phi)\dot{x}+x=F(x,\phi),
\end{equation}
the solution of which may be analyzed via Poincare's theory. Here, the
nonlinear force $F(x,\phi)$ and friction coefficient $\eta(x,\phi)$, which in what follows we refer to as a pumping coefficient,
generated by interaction with the nonequilibrium electronic environment
take the form,
\begin{eqnarray}
  &&F(x,\phi)= \kappa\xi\frac{\tilde{\Delta}_d}{2\tilde{\mathcal{D}}^2}\left[ \sin{\phi}-\tilde{\varepsilon}_d\sinh{\left(2x\right)}\right],\label{fxphi}\\
  &&\eta(x,\phi)=\kappa\alpha\xi\frac{\tilde{\Delta}_d}{2\tilde{\mathcal{D}}^6}\left[ \tilde{\Delta}_d^2\sinh{(2x)}\left(\tilde{\mathcal{D}}^2+4\right)
  \left\{ \sin{\phi}-\tilde{\varepsilon}_d\sinh{(2x)}\right\}+8\tilde{\varepsilon}_d\tilde{\mathcal{D}}^2\left\{\sin^2{(\phi/2)}+\sinh^2{x}\right\}\right].\label{etaxphi}
\end{eqnarray}

Here
$\tilde{\mathcal{D}}^{2}\equiv\mathcal{D}^2(x,\phi)/\Gamma_n^2=\tilde{\Delta}_d^2\left[ \sinh^2{x}+\cos^2{(\phi/2)}\right]+\tilde{\varepsilon}^{2}_{d}+1$.

\section{Self-sustained oscillations}

In order to find the stationary solutions $x_{c}(t)$ in
Eq.~(\ref{xEqsol}), it is natural to use the smallness of the parameter
$\xi$ and look for such solutions~\cite{bogoliubov} in the form
$x_{c}(t)=x_{st}+\sqrt{A}\sin\left(t+\varphi(t)\right)+\mathcal{O}(\xi)$,
where $x_{st},\dot{A}(t),\dot{\varphi}(t)\sim \xi$. Then, with the accepted
accuracy $\xi$, one can get the following equations:
\begin{eqnarray}
      &&\dot{A}=  A\bar{\eta}(A,\phi), \label{dotA} \\
      && \dot{\varphi}=- A^{-1/2}\bar{F}(A,\phi),\label{Ephi}
\end{eqnarray}\label{EA}
where
\begin{eqnarray}
&&\bar{\eta}(A,\phi)=\kappa(\pi)^{-1}\int_{0}^{2\pi}d\psi
\cos^2{(\psi)}
\eta\left(\sqrt{A}\sin\psi,\phi\right)\equiv\kappa\xi\alpha
W(A,\phi),\label{W}\\
&&\bar{F}(A,\phi)=\kappa(2\pi)^{-1}\int_{0}^{2\pi}d\psi\sin{(\psi)}F\left(\sqrt{A}\sin\psi,\phi\right).
\end{eqnarray}
The pumping coefficient $\bar{\eta}(A,\phi)$ has an obvious physical
meaning: it gives the ratio between the energy supplied into the
mechanical degree of freedom for one period of mechanical
oscillation with amplitude $\sqrt{A}$ and the total mechanical
energy.

\begin{figure}
    \centering
    \begin{subfigure}[t]{0.4\columnwidth}
        \centering
        \includegraphics[width=\columnwidth, keepaspectratio]
        {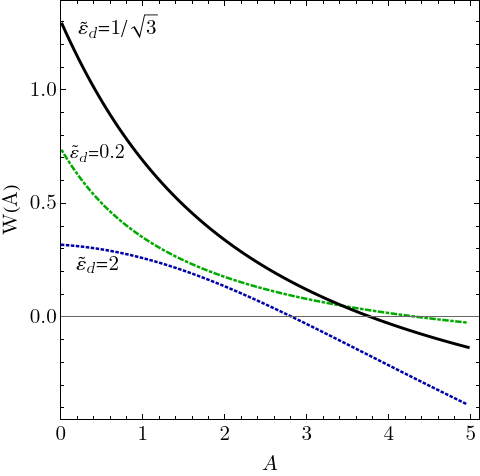}
        \caption{}\label{fig:subfig3a}
    \end{subfigure}
    \quad
    \begin{subfigure}[t]{0.4\columnwidth}
        \centering
        \includegraphics[width=\columnwidth, keepaspectratio]
        {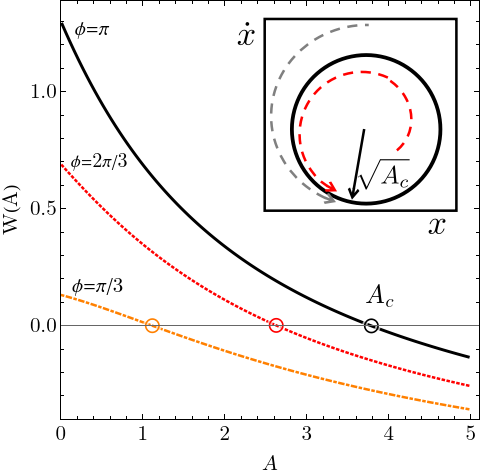}
        \caption{}\label{fig:subfig3b}
    \end{subfigure}
    \caption{Plots of the function $W(A)$, proportional to the pumping coefficient $\bar{\eta}(A,\phi)$, for different values of
    (a) the relative position of the dot energy level $\tilde{\varepsilon}_d=0.2; 1/\sqrt{3}; 2$ for $\phi=\pi$, and
    (b) the superconducting phase difference $\phi=\pi/3, 2\pi/3, \pi$ for $\tilde{\varepsilon}_d=1/\sqrt{3}$.
     The zeroes of the functions correspond to the amplitude of the limiting cycle [see the inset in (b)], which strongly depends on the superconducting
     phase difference and reaches its maximum at $\phi=\pi$. The other parameters are $\tilde{\Delta}_d=1, \kappa=+1$. }\label{fig:WA}
\end{figure}

\begin{figure}
    \centering
    \begin{subfigure}[t]{0.4\columnwidth}
        \centering
        \includegraphics[width=\columnwidth, keepaspectratio]
        {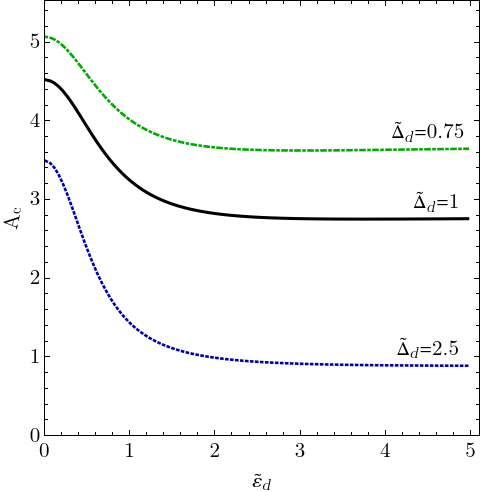}
        \caption{}\label{fig:subfig4a}
    \end{subfigure}
    \quad
    \begin{subfigure}[t]{0.4\columnwidth}
        \centering
        \includegraphics[width=\columnwidth, keepaspectratio]
        {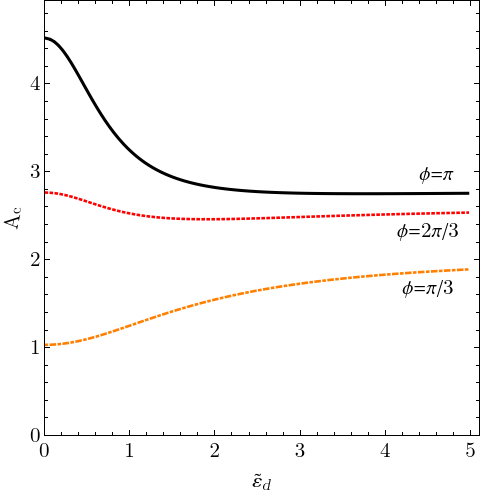}
        \caption{}\label{fig:subfig4b}
    \end{subfigure}
    \caption{Dependencies of $A_c$ of the limiting cycles on the relative position of the dot energy level $\tilde{\varepsilon}_d$ (counted from the Fermi energy) for different values of (a) $\tilde{\Delta}_d=0.75; 1; 2.5$ for $\phi=\pi$, and (b) the superconducting phase difference $\phi=\pi/3, 2\pi/3; \pi$ for $\tilde{\Delta}_d=1$.}\label{fig:Ac}
\end{figure}

It is evident from Eq.~(\ref{dotA}) that stationary regimes $\dot{A}=0$
are given by equations $A=0$ ($x(t)=x_{st}$) and
$\bar{\eta}(A,\phi)=0$. The first one is a static state of the
nanotube, and the second one corresponds to periodic oscillations with the 
amplitude $\sqrt{A_{c}}$, where $W(A_{c},\phi)=0$. The static regime is stable when $ \bar {\eta} (0, \phi)
<0 $ and unstable otherwise. The stability of the periodic solution
is defined by the sign of the derivative
$\partial_A{\bar{\eta}(A,\phi)}|_{A=A_{c}}$: if it is negative (positive), then the
periodic regime is stable (unstable). Analyzing Eqs.~(\ref{xEqsol}) and (\ref{W}), one can conclude that the
pumping coefficient $\bar{\eta}(A,\phi) \propto \kappa$ is an odd
function of $\varepsilon_{d}$ and takes the following limit values,
\begin{eqnarray}
   && \bar{\eta}(0,\phi)=\kappa\alpha\xi W(0,\phi)=+\kappa\alpha\xi\frac{4\tilde{\varepsilon}_d\tilde{\Delta}_d}{\tilde{\mathcal{D}}^4(0,\phi)}\sin^2{\left(\frac{\phi}{2}\right)},\label{etaphi}\\
  &&\eta(A\rightarrow\infty,\phi)= \kappa\alpha\xi W(A \rightarrow\infty,\phi)=-\kappa\alpha\xi\frac{\tilde{\varepsilon}_d}{2\tilde{\Delta}_d},\label{etaphiInfty}
\end{eqnarray}
from which follows that at $\phi\neq 0$, the solution of the equation
$W(A_{c},\phi)=0$, corresponding to the stationary periodic regime, exists at
any values of the other parameters. The case when $\phi=0$ is very
unstable with respect to the small asymmetry parameter $|a|\ll 1$
[see below Eq.~(\ref{Hts})] and will be analyzed elsewhere. The function
$W(A)$ and $A_c$ at different $\phi\geq 1$ and $\tilde{\varepsilon}_{d}>0$ are
presented in Figs.~\ref{fig:WA} and \ref{fig:Ac}.

From  Eq.~(\ref{etaphi}), it follows that if $\kappa \varepsilon_{d}>0$
and $\phi\neq 0$, the static mechanical state $x=x_{st}\ll1$ is
unstable with respect to the appearance of bending oscillations with
amplitudes that exponentially increase in time with the increment
$\gamma=\kappa\alpha\xi W(0,\phi)$. The latter takes its maximum at
$\phi=\pi$ for the fixed values of all other parameters (notice that
$x_{st}(\pi)=0$). However, the increase saturates at the amplitude
$\sqrt{A_{c}}$, resulting in self-sustained oscillations at this
amplitude. It should be noted that the amplitude saturation in the system
under consideration is a completely internal effect and still takes place
when the friction caused by interaction with a
thermodynamic environment is zero. A "self-saturation" effect was
also reported in~\cite{ilinskaya} where a special magnetic NEM system was
considered.

\section{Electric current}

\begin{figure}
\includegraphics[width=.45\columnwidth]{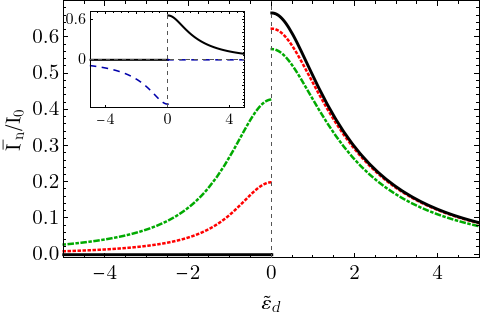}
\caption{Dependencies of the dc electric current $\bar{I}_n$ (normalized to $I_0=e\Gamma_n$) on the relative position of the QD energy level $\tilde{\varepsilon}_d$ for different values of superconducting phase difference $\phi=\pi/3$ (green dot-dashed curve), $2\pi/3$ (red dotted), and $\pi$ (black solid) for $\tilde{\Delta}_d=1$ and $\kappa=+1$. The maximum effect occurs at $\phi=\pi$ when the dc current is absent in the static regime, while it is close to the maximum one in the stable stationary regime of the self-sustained oscillations.
Inset: Dependencies of the time-averaged electric current for $\kappa=+1$ (black solid curve, associated with the same one in the main plot) and for $\kappa=-1$ when the bias voltage is applied in the opposite direction (blue dashed curve), representing a diode-like behaviour of the current.
}\label{fig:fig5}
\end{figure}

The self-sustained oscillations considered above have a very specific
transport signature. This raises the possibility of detecting the
mechanical instability through electric current measurement. To
explore such a possibility, let us consider the electric current through
the system $I_{n}$, determined in a standard way,
$I_n=e\kappa\text{Tr}\left\{\hat{\dot{N}}\hat{\rho}\right\}$, where
$\hat{\dot{N}}= i[\hat{H},\hat{N}]$ and $\hat{N}$ is the operator of
the number of electrons in the normal electrode. In the high bias voltage
regime at $\kappa=+1$, where electron tunneling from the QD to the normal leads is forbidden, an expression for $I_{n}$ can
be easily obtained by analyzing Fig.~\ref{fig:fig2}. From
those diagrams, one can see that a decrease in the number of electrons
in the normal electrode is defined by two different processes. The first
one is the tunneling of an electron with spin up or down into the empty
dot. The rate of this process is $2\Gamma_{n}\rho_{0}$, where
$\rho_{0}=\left( R_{0}+R_{3}\right) /2$ is the probability that the dot is empty.
The second one is the tunneling of an electron into the dot occupied by
 a single electron with spin up or down. The rate of this process is
$2\Gamma_{n}\rho_{e}$. Taking into account the normalization condition
$2\rho_{e}+R_{0}=1$, and using a similar speculation for $eV_b<0$, one gets
from Eq.~(\ref{Rexpand}) the following equation for $I_{n}$,
\begin{equation}\label{Icurrent}
    I_n(t)=\kappa I_{0}\left(1+\kappa R_3\right)=
    \kappa I_{0}\left[\frac{|\Delta_{d}(x,\phi)|^{2}}{\mathcal{D}^{2}(x,\phi)}+ \alpha \dot{x}f(x)+\mathcal{O}(\alpha^{2})\right],
\end{equation}

where $I_0=e\Gamma_n$. In the stationary regime
corresponding to the generation of self-sustained oscillations with
amplitude $\sqrt{A_{c}}$, the average electric current
$\bar{I}_{n}$ is defined by the equation
\begin{equation}
    {\bar{I}}_{n}(\kappa,\varepsilon_{d})=\kappa I_0\left[\frac{\Delta_d^2\cos^{2}{(\phi/2})}{\Delta_d^2\cos^{2}(\phi/2)+\Gamma_n^2+\varepsilon_d^2}
+ \theta(\kappa\varepsilon_{d})\delta\bar{I}(A_{c})\right],
\end{equation}\label{Iav}
 where the first term corresponds to the static dc current, which crucially depends on the superconducting
 phase difference $\phi$. In particular, the first term is equal to zero at $\phi=\pi$, in contrast to the second term,
\begin{equation}\label{deltaI}
    \delta\bar{I}(A_{c})=\dfrac{1}{2\pi}\dfrac{\Delta_d^2 (\Gamma_n^2+\varepsilon_d^2)}{\mathcal{D}^2(0,\phi)}\int_0^{2\pi}d\psi \dfrac{\sinh^2{\left(\sqrt{A_c}\sin{\psi}\right)}}{\mathcal{D}^2\left( \sqrt{A_c}\sin{\psi},\phi\right)}>0,
\end{equation}
which emerges exclusively due to the self-sustained oscillations and
equals zero if the static state is stable, as indicated by the Heaviside
step function $\theta(\kappa\varepsilon_d)$. Plots of
$\bar{I}_{n}$ as a function of $\tilde{\varepsilon}_{d}$ at
$\tilde{\Delta}_{d}=1$ and $\phi=\pi/3, 2\pi/3, \pi$ are presented in
Fig.~\ref{fig:fig5}. These graphs show that the nanomechanical
instability discussed in this article leads to the emergence of
significant diode and transistor effects. The effects are most
pronounced at $\phi=\pi$ when in the static regime $A_{c}=0$ where the Cooper pair exchange between the dot and the superconducting leads is completely blocked. In such a situation, a jump in the average current
from zero to a finite value $\sim I_{0}$ (or vice-versa) occurs if the direction
of the bias voltage changes (diode effect) or if the position of the level $\varepsilon_{d}$ controlled by the gate voltage passes zero
(transistor effect). Note that the discontinuity of the average current
as a function of $\varepsilon_{d}$ must be treated to the accuracy
$\xi$ accepted in this paper.

\section{Conclusions}

In summary, we considered a nanoelectromechanical system comprising a carbon nanotube suspended above a trench in a normal
metal electrode that undergoes bending vibrations in the gap
between two superconductors. The nanotube was treated as a movable
single-level quantum dot in which the superconducting order
parameter is induced as a result of Cooper pair exchange with the
superconductors. The latter essentially depends on the bending of
the nanotube and the phase difference between the superconductors. We
have shown that in such a system, the static, straight configuration
of the nanotube is unstable regarding the occurrence of
self-sustained bending vibrations in a wide range of parameters if a
bias voltage is applied between the normal and superconducting leads. It was demonstrated that the occurrence of this
instability crucially depends on the direction of the bias voltage
and the relative position of the QD level. This makes it possible to
govern the operating mode of the system by changing the bias and
gate voltages. We have also shown that the appearance of
self-sustained mechanical vibrations strongly affects the dc current
through the system, leading to transistor and diode effects. The
latter can be used for the direct experimental observation of the
predicted phenomena.

\section*{Acknowledgements}
O.M.B. thanks A.V.~Parafilo and O.A.~Ilinskaya for helpful discussions.
S.I.K. acknowledges the financial support from the NAS of Ukraine (grant F 26-4). This work was supported by IBS-R024-D1.

\end{document}